\documentclass[preprint,12pt]{elsarticle}
\newcommand\beq{\begin{eqnarray}}
\newcommand\eeq{\end{eqnarray}}
\newcommand\bq{\begin{equation}}
\newcommand\eq{\end{equation}}
\usepackage{graphicx}
\usepackage{amssymb}

\journal{Nuclear Physics B}

\begin{document}

\begin{frontmatter}



\title{CP violation in scattering of neutrinos on  polarized proton target}
\author{W. Sobk\'ow\corref{cor}}
 \ead{sobkow@sunflower.ift.uni.wroc.pl}
 \cortext[cor]{Corresponding author}
\address{Institute of
Theoretical Physics, University of Wroc\l{}aw, Pl. M. Born 9,
\\PL-50-204~Wroc\l{}aw, Poland} 

\begin{abstract}
In this paper, we analyze the elastic scattering of the muon neutrino beam on
the polarized proton target (PPT), and predict how the existence of CP-violating phase between
the complex vector V and axial A couplings of the left-chirality 
neutrinos affects the azimuthal dependence of the differential cross section. 
The neutrinos are assumed to be Dirac fermions with non-zero mass. 
We show that the azimuthal asymmetry of recoil protons does not 
depend on the neutrino mass and does not vanish  even if
$\beta_{VA} = 0$. The CP-breaking phase $\beta_{VA}$ could be
detected by measuring the maximal asymmetry of the differential cross section. We also indicate the possibility of using the PPT to distinguish the detector background from the neutrino interactions. 
Our analysis is model-independent and  the results are presented in a limit of infinitesimally small neutrino mass. 
 
 \end{abstract}

\begin{keyword}
neutrino nucleon scattering \sep polarized proton target 


\end{keyword}

\end{frontmatter}


\section{\label{intro} Introduction}
Neutrino-nucleon scatterings (NNS) play a crucial role in testing the predictions of the standard model (SM) of electroweak interactions \cite{Glashow,Wein,Salam}. 
These reactions are also  very important for the neutrino oscillation experiments and understanding the dynamics of the core-collapse supernovae \cite{Balantekin}. In addition, the NNS  contribute  to the energy transfer from the accretion-disk neutrinos to the nucleons 
\cite{Raffert,Kneller} and are essential to the terrestrial neutrino observation \cite{Vogel,Beacom,Jesus}. A measurement of the neutrino-nucleon neutral current elastic scattering cross section  allows to search for contribution of the strange sea quarks to the nucleon spin  (SciBooNE experiment). 
The experiments with the NNS can be carried out  using the intense, pulsed spallation sources, beta-beams facilities, superbeams and neutrino factories  \cite{Avignone,Scholberg,CVolpe,Geer}.  Various applications of the neutrino beams in area of the NNS and SM tests have been discussed in  literature \cite{Serreau,Mc,Volpe,McLau,JHJesus}.  \\ 
Although the SM  agrees with the experimental data up to available energies, there are numerous theoretical reasons for which it can not be viewed as a ultimate theory. The SM does not clarify why parity is
 violated in the weak interaction and what is the mechanism behind
this violation.  The another fundamental problem  is impossibility of explaining  the observed baryon asymmetry of universe 
\cite{barion} through  a single CP-violating phase of the Cabibbo-Kobayashi-Maskawa quark-mixing matrix (CKM) \cite{Kobayashi}. Presently the CP violation is observed only in the decays of neutral K- and B-mesons \cite{CP,Todd}. 
Up to now  there is no direct evidence of the CP violation in the
leptonic and semileptonic processes. 
However, the future superbeam and neutrino factory experiments 
 aim at the measurement of  the CP-violating effects in the lepton
sector, where both neutrino and antineutrino oscillation might be observed.  The search of the CP violation in the semileptonic reactions involves mainly the precise measurements of T-odd triple-correlations between the outgoing particles momenta and initial nucleon (nucleus) polarization \cite{neutron}. All the results are still compatible with the CP conservation scenario. \\ 
In such a situation, it is meaningful to look for new tools essential for   testing the CP symmetry breaking in the neutral and charged weak interactions. 
 It is necessary to point out that  so far no measurements with the (quasi)elastic scattering of neutrino beam on the polarized nucleon (proton, neutron) targets (PNT) have been carried out. One should note that various kinds of the PPT have been used for the nucleon-nucleon elastic scattering and transmission experiments for 15 years, e.g. SATURNE polarized target \cite{Saturne}.  The SATURNE Collaboration experience  should be used in studies on the future PNT.  \\
In this paper, we focus on the elastic scattering of the muon neutrino beam off the PPT, and  consider how the existence of non-zero CP-violating  phase between
the complex vector V and axial A couplings of the left-chirality neutrinos affects the azimuthal dependence of the differential
cross section. Next, we show how the recoil proton energy spectrum depends on the initial  polarization of proton target. \\
Our considerations are model-independent and  the calculations are made in the limit of
infinitesimally small neutrino mass.  We use the
system of natural units with $\hbar=c=1$, Dirac-Pauli representation of the
$\gamma$-matrices and the $(+, -, -, -)$ metric \cite{Greiner}.

\section{\label{PPT} Scattering of left-chirality muon neutrinos on polarized protons}

\begin{figure}
\begin{center}
\includegraphics*[scale=.4]{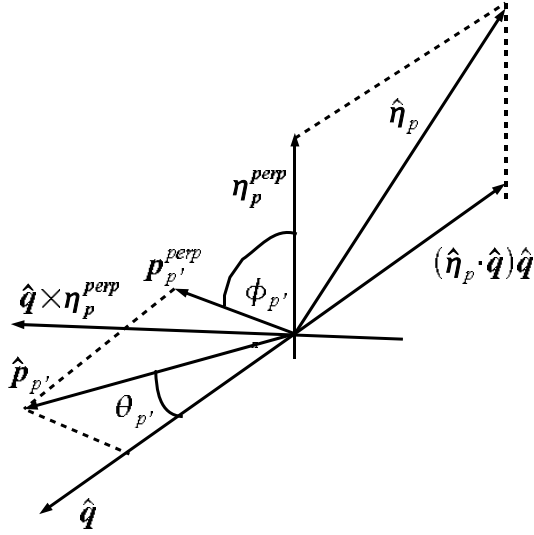}
\end{center}
\caption{Kinematics and  reaction plane for the $\nu_\mu p$
scattering.} \label{rp}
\end{figure}

We consider the incoming muon
neutrino beam, which consists only of the left-chirality and
longitudinally polarized neutrinos. We assume
that these neutrinos are detected in the standard $V-A$ neutral current weak
interactions with the PPT and both the recoil proton scattering angle
$\theta_p'$ and the azimuthal angle of outgoing proton momentum
$\phi_p'$, shown in Fig. \ref{rp}, are measured with a good angular resolution. 
 As our analysis is carried out in the limit of vanishing $\nu_\mu $ mass,  left-chirality $\nu_\mu $ has 
negative helicity. 
We admit  the CP violation (equivalent the time reversal violation), so the
amplitude includes the complex coupling constants denoted as
 $|g_V^L| \,  e ^{i \beta_V^L}$, $g_A^L = |g_A^L| \,  e ^{i \beta_A^L}$ respectively to the initial neutrino of left-chirality:
  \beq M_{\nu_{\mu} p} &=&
\frac{G_{F}}{\sqrt{2}}\Bigg\{g_{V}^{L}(\overline{u}_{p'}\gamma^{\alpha}u_{p})
(\overline{u}_{\nu_{\mu'}} \gamma_{\alpha}(1 -
\gamma_{5})u_{\nu_{\mu}})  \\ 
&& \mbox{} +g_{A}^{L} (\overline{u}_{p'}\gamma^{5}\gamma^{\alpha}u_{p})
(\overline{u}_{\nu_{\mu'}} \gamma_{5}\gamma_{\alpha}(1 -
\gamma_{5})u_{\nu_{\mu}}) \Bigg\}, \nonumber\eeq  
where $ u_{p}$ and  $\overline{u}_{p'}$
$(u_{\nu_{\mu}}\;$ and $\; \overline{u}_{\nu_{\mu'}})$ are the
Dirac bispinors of the initial and final proton (neutrino)
respectively. $G_{F} = 1.1663788(7)\times
10^{-5}\,\mbox{GeV}^{-2} (0.6 ppm)$ (MuLan Collab.) \cite{Mulan} is the Fermi constant. We point out that the induced weak coupling of the left-chirality  neutrino ($g_{M}^L$- the weak
magnetism) enters additively  the 
$g_{V}^{L} $ coupling, so it is omitted because its presence does not change qualitatively the conclusions concerning the  CP violation. 

By $\mbox{\boldmath $\hat{\eta}_{\overline{\nu}}$}$ and 
$(\mbox{\boldmath$\hat{\eta}_{\overline{\nu}}$}\cdot\hat{\bf q}){\bf\hat{q}}$,
 we denote the unit
polarization vector and its longitudinal component of the
incoming $\nu_{\mu}$ in its rest system, respectively. 
Then $\mbox{\boldmath $\hat{\eta}_{\nu}$}\cdot\hat{\bf q} = -1 $
is the longitudinal polarization of the $\nu_{\mu}$. ${\bf \hat{q}}$ is the incoming muon neutrino
LAB momentum unit vector; ${\bf
\hat{p}_{p'}}$ is the unit vector of the outgoing proton momentum; $ \mbox{\boldmath
$\hat{\eta}_{p}$}$ is the unit 3-vector of the initial proton 
polarization in its rest frame. The measurement of the
azimuthal angle of outgoing proton momentum $\phi_{p'}$ is only
possible when the proton target polarization is known. The $\phi_{p'}$  is
measured with respect to the transverse component of the initial
proton polarization {\boldmath $\eta_{p} ^{\perp}$}.  $\theta_{p'}$ is the
polar angle between the  $\hat{\bf p}_{p'}$ and the  $\hat{\bf q}$ (recoil proton scattering
angle). As it is well known, the vector $\mbox{\boldmath $\hat{\eta}_{p}$}$  can
be expressed, with respect to the $\hat{\bf q}$, as a sum of the longitudinal
component of the proton polarization
$(\mbox{\boldmath$\hat{\eta}_{p}$}\cdot\hat{\bf q}){\bf \hat{q}} $
 and transverse component of the proton polarization 
$\mbox{\boldmath $\eta_{p} ^{\perp}$} $, that is defined as
$ \mbox{\boldmath $\eta_{p}^{\perp}$} =
\mbox{\boldmath$\hat{\eta}_{p}$}-
(\mbox{\boldmath$\hat{\eta}_{p}$}\cdot\hat{\bf q}) {\bf \hat{q}}$, Fig. 1.

\section{Azimuthal distribution of recoil protons} 
The formula for the azimuthal distribution of the scattered protons  with $\mbox{\boldmath $\hat{\eta}_{p}$} \perp {\bf \hat{ q}} $ is of the form: 
\beq \label{SMcomplex} \lefteqn{\left(\frac{d^2 \sigma}{d y d
\phi_{p'}}\right)_{(V A)}  = \frac{E_\nu
m_{p}}{4\pi^2}\frac{G_{F}^{2}}{2}(1-\mbox{\boldmath
$\hat{\eta}_{\nu}$}\cdot\hat{\bf q})}\\
 && \cdot \Bigg\{ |g_{A}^{L}|^2 \left[-
\mbox{\boldmath $\hat{\eta}_{p}\cdot
\hat{p}_{p'}$}\sqrt{\frac{2m_p}{E_\nu}+y} (\sqrt{y^{3}}-2\sqrt{y})
+ \frac{m_p}{E_\nu}y + (y-2)y +2 \right] \nonumber \\
&& \mbox{} + |g_{V}^{L}|^2 \left[ y^2 - \mbox{\boldmath
$\hat{\eta}_{p}\cdot
\hat{p}_{p'}$}\sqrt{y^{3}}\sqrt{\frac{2m_p}{E_\nu}+y} -
y(\frac{m_p}{E_\nu} + 2) +2 \right]\nonumber \\
&& \mbox{} +
2 Im(g_{V}^L g_A^{L*}) \, {\bf \hat{q}} \cdot
(\mbox{\boldmath $\hat{\eta}_{p} \times \hat{p}_{p'} $})\sqrt{y(\frac{2m_p}{E_\nu}+y)}\nonumber\\
&& \mbox{} +
 2Re(g_{V}^L g_A^{L*}) \left[
\mbox{\boldmath $\hat{\eta}_{p}\cdot
\hat{p}_{p'}$}(y-1)\sqrt{y(\frac{2m_p}{E_\nu}+y)} + (2-y)y \right]
\Bigg\}. \nonumber \eeq
 The variable $y $ is the ratio of the kinetic energy
of the recoil proton $T_{p} $ to the incoming neutrino energy
$E_{\nu} $:
\beq
y\equiv\frac{T_{p}}{E_{\nu}}=\frac{m_{p}}{E_{\nu}}\frac{2cos^{2}\theta_{p'}}
{(1+\frac{m_{p}}{E_{\nu}})^{2}-cos^{2}\theta_{p'}}. \eeq
It varies from $0 $ to $2/(2+m_p/E_\nu)$, where $m_{p}$ is  the proton mass. 
\par After the simplification of the above formula, Eq. (\ref{SMcomplex}),
we obtain  the new form: 
\beq \lefteqn{\left(\frac{d^2 \sigma}{d y d \phi_{p'}}\right)_{(V A)} =
\frac{E_\nu m_{p}}{4\pi^2}\frac{G_{F}^{2}}{2}(1-\mbox{\boldmath
$\hat{\eta}_{\nu}$}\cdot\hat{\bf q}) \Bigg\{  |\mbox{\boldmath
$\eta_{p}^{\perp}$}|\sqrt{\frac{m_p}{E_\nu}y[2-y(2+\frac{m_p}{E_\nu})]}}\nonumber\\
&& \mbox{} \cdot \bigg[ cos(\phi_{p'}) \left(2|g_V^L||g_A^L|
cos(\beta_{VA})y  + (2-y)|g_{A}^{L}|^{2} - y|g_{V}^{L}|^2 \right)\nonumber\\
&& \mbox{} - 2|g_V^L||g_A^L|cos(\phi_{p'} + \beta_{VA}) \bigg]  
 +  \bigg[\left(|g_{V}^{L}|^2 + |g_{A}^{L}|^2\right)
\left( y^2 -2y +2 \right) \nonumber\\
 && \mbox{} + 2|g_V^L||g_A^L| cos(\beta_{VA})y(2-y)
- \frac{m_p}{E_\nu}y \left(|g_{V}^{L}|^2 -
|g_{A}^{L}|^{2}\right)\bigg] \Bigg\}. \eeq

 It can be noticed that the interference terms between the
standard $g_{V, A}^{L}$  couplings depend on the value of the relative phase
$\beta _{VA} = \beta _{V}^L - \beta _{A}^L$  and are independent of the muon neutrino mass. In the interference part we have angular correlations with the
transverse component of the PPT 
$\mbox{\boldmath $\eta_{p}^{\perp}$}$, both T-odd and T-even. We see that the azimuthal asymmetry of the recoil protons does not vanish  even if $\beta_{VA} = 0$. The
CP-violating phase enters the differential cross section and changes the angle
at which the number of the recoil protons will be maximal 
$(\phi_{p'}^{max})$.  For $\beta_{VA} = \frac{\pi}{2}$ and
$|g_V^L| = 1, |g_A^L| = 1.26$,  this angle  is quite large
$\phi_{p'}^{max} \simeq \frac{\pi}{3}$, see Fig. 2, 3 (solid lines). The azimuthal dependence is illustrated for two assumed values of the muon neutrino energies, i. e.  $E_{\nu}=50 MeV$ and $E_{\nu}= 1 GeV$. We calculate the extreme of a function stying at $|\mbox{\boldmath $\eta_{p}^{\perp}$}|$ and  get  $y_{max} \simeq 0.048$ for   $E_{\nu}=50 MeV$,  and  $y_{max} \simeq 0.34$ for   $E_{\nu}= 1 GeV$, respectively. In
the case of pure axial-vector $g_A^L$ coupling (short-dashed line) we have different
azimuthal dependence of the cross section ($\phi_{p'}^{max} = 0$) 
than in the case of pure  vector $g_V^L$ coupling (long-dashed line) 
$(\phi_{p'}^{max} = \pm \pi)$. The dotted lines in Fig. 2, 3 show the expected  number of  scattered protons for the CP-symmetric scenario. 
\begin{figure}
\begin{center}
\includegraphics*[scale=.7]{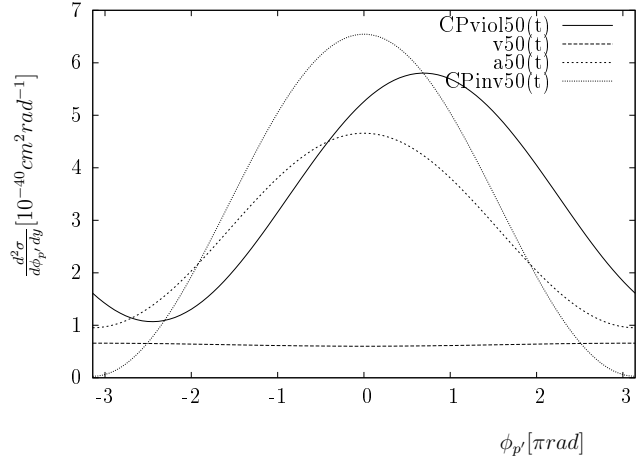}
\end{center}
\caption{Plot of the $\frac{d^{2} \sigma}{ d y d \phi_p'}_{(V A)}$ as a
function of the azimuthal angle $\phi_{p'}$ for the
$(\nu_{\mu}p)$ scattering, when $E_{\nu} = 50 \mbox{ MeV}$, $y =
0.048$, $\mbox{\boldmath$\hat{\eta}_{\nu}$}\cdot\hat{\bf q} = - 1$ 
and $|\mbox{\boldmath $\eta_{p}^{\perp}$}| = 1$: 
a) CP
violation with   $|g_V^L| =1, 
|g_A^L| = 1.26$  and $\beta_{VA} =
\frac{\pi}{2}$ (solid line),  b) $g_A^L = 0, g_V^L = 1$ (long-dashed line),   $g_V^L =0, g_A^L = -1.26$ (short-dashed line), d) CP
conservation with $|g_V^L| =1, 
|g_A^L| = 1.26$ and  $\beta_{VA} = \pi$ (dotted line).}  \label{aaCPviol50}
\end{figure}
\begin{figure}
\begin{center}
\includegraphics*[scale=.7]{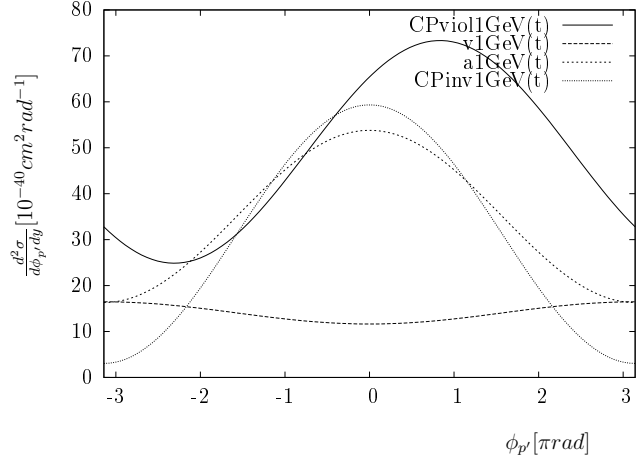}
\end{center}
\caption{Plot of the $\frac{d^{2} \sigma}{ d y d \phi_p'}_{(V A)}$ as a
function of the azimuthal angle $\phi_{p'}$ for the
$(\nu_{\mu}p)$ scattering, when  $E_{\nu} = 1 \mbox{ GeV}$, $y =
0.34$, $\mbox{\boldmath$\hat{\eta}_{\nu}$}\cdot\hat{\bf q} = - 1$ 
and $|\mbox{\boldmath $\eta_{p}^{\perp}$}| = 1$: 
a) CP
violation with   $|g_V^L| =1, 
|g_A^L| = 1.26$  and $\beta_{VA} =
\frac{\pi}{2}$,  b) $g_A^L = 0, g_V^L = 1$ (long-dashed line),   $g_V^L =0, g_A^L = -1.26$ (short-dashed line), d) CP
conservation with $|g_V^L| =1, 
|g_A^L| = 1.26$ and  $\beta_{VA} = \pi$ (dotted line).}  \label{aaCPviol1GeV}
\end{figure}

\section{Spectrum of recoil protons}

When the outgoing proton direction
is not observed,  the formula for the laboratory differential
cross section, Eq. \ref{SMcomplex}, has to be integrated over the azimuthal
angle $\phi_{e'}$ of the recoil proton momentum. 
Finally, one obtains  for  $\mbox{\boldmath $\hat{\eta}_{p}$} \not\perp {\bf \hat{ q}}$: 

\beq  \label{spproton} \left(\frac{d \sigma}{d y }\right)_{(V, A)}
&=& \frac{E_{\nu}m_{p}}{2\pi}\frac{G_{F}^{2}}{2}(1-\mbox{\boldmath
$\hat{\eta}_{\nu}$}\cdot\hat{\bf q})\Bigg\{\left|g_{V}^{L} +
g_{A}^{L}\right|^{2}(1 + \mbox{\boldmath
$\hat{\eta}_{p}$}\cdot{\bf\hat{q}}) \\
&& \mbox{} + \left|g_{V}^{L} -
g_{A}^{L}\right|^{2}\left[1 - (\mbox{\boldmath
$\hat{\eta}_{p}$}\cdot {\bf \hat{q}})\left(1-
\frac{m_{p}}{E_{\nu}}\frac{y}{(1-y)}\right)\right](1-y)^{2}
\nonumber\\
& & \mbox{} - \left[\left|g_{V}^{L}\right|^{2} -
\left|g_{A}^{L}\right|^{2}\right] (1 +
\mbox{\boldmath $\hat{\eta}_{p}$}\cdot{\bf\hat{q}})\frac{m_{p}}{E_{\nu}}y\Bigg\}\nonumber. 
\eeq
We see that the proton energy spectrum  contains only the T-even contributions $\mbox{\boldmath
$\hat{\eta}_{p}$}\cdot {\bf \hat{q}}$ contrary to the azimuthal recoil proton distribution, where  the T-odd correlation occurs. The measurement of the projection of the proton polarization vector parallel to the ${\bf \hat{ q}}$ is only possible when the polarization of the proton target is known. For $E_{\nu}= 50 MeV$ and  $E_{\nu}= 1 GeV$,  we present three interesting cases for $\mbox{\boldmath
$\hat{\eta}_{p}$}\cdot {\bf \hat{q}}=-1, 0, 1$, shown in Fig. 4, 5.   
We assume that the detector energy threshold is $T_{p'}^{th}=1 keV$, so $y \in [0.001, 0.096]$  for $E_{\nu}= 50 MeV$ and   $y \in [0.001, 0.68]$ for   $E_{\nu}= 1 GeV$, respectively. It can be noticed that the significant growth of the differential cross section  for  $\mbox{\boldmath $\hat{\eta}_{p}$}\cdot {\bf \hat{q}}= 1$ with  $E_{\nu}= 50 MeV$ (solid  line, Fig. 4) is seen for the proton kinetic energy near by the detector threshold.  For 
 $\mbox{\boldmath $\hat{\eta}_{p}$}\cdot {\bf \hat{q}}=-1$ (long-dashed line, Fig. 4), the number of recoil protons strongly decreases with respect to the unpolarized case  $\mbox{\boldmath $\hat{\eta}_{p}$}\cdot {\bf \hat{q}}=0$ (short-dashed line) and is also seen nearly the detector threshold.   In the case of the GeV  $\nu_{\mu} p$ scattering, the increase (and decrease) of the differential cross section for  
 $\mbox{\boldmath $\hat{\eta}_{p}$}\cdot {\bf \hat{q}}= 1$ (and for $\mbox{\boldmath $\hat{\eta}_{p}$}\cdot {\bf \hat{q}}= -1$) takes place in the whole $y$ range from the detector threshold to the kinetic maximum, see Fig. 5 with the solid line (and long-dashed line).  
 The dotted lines, shown in the Fig. 4, 5,  represent the energy spectrum of protons according to the SM prediction for the unpolarized proton target. 
 \par One ought point out that the neutrino beam scattering on the PPT can also be used to distinguish the isotropic background from the proper neutrino interactions, because the incoming left-chirality neutrinos mainly interact with the left-chirality protons. If the initial neutrino beam is scattered off the right-chirality protons, the event number is suppressed, but the detector background stays the same.  
 \begin{figure}
\begin{center}
\includegraphics*[scale=.7]{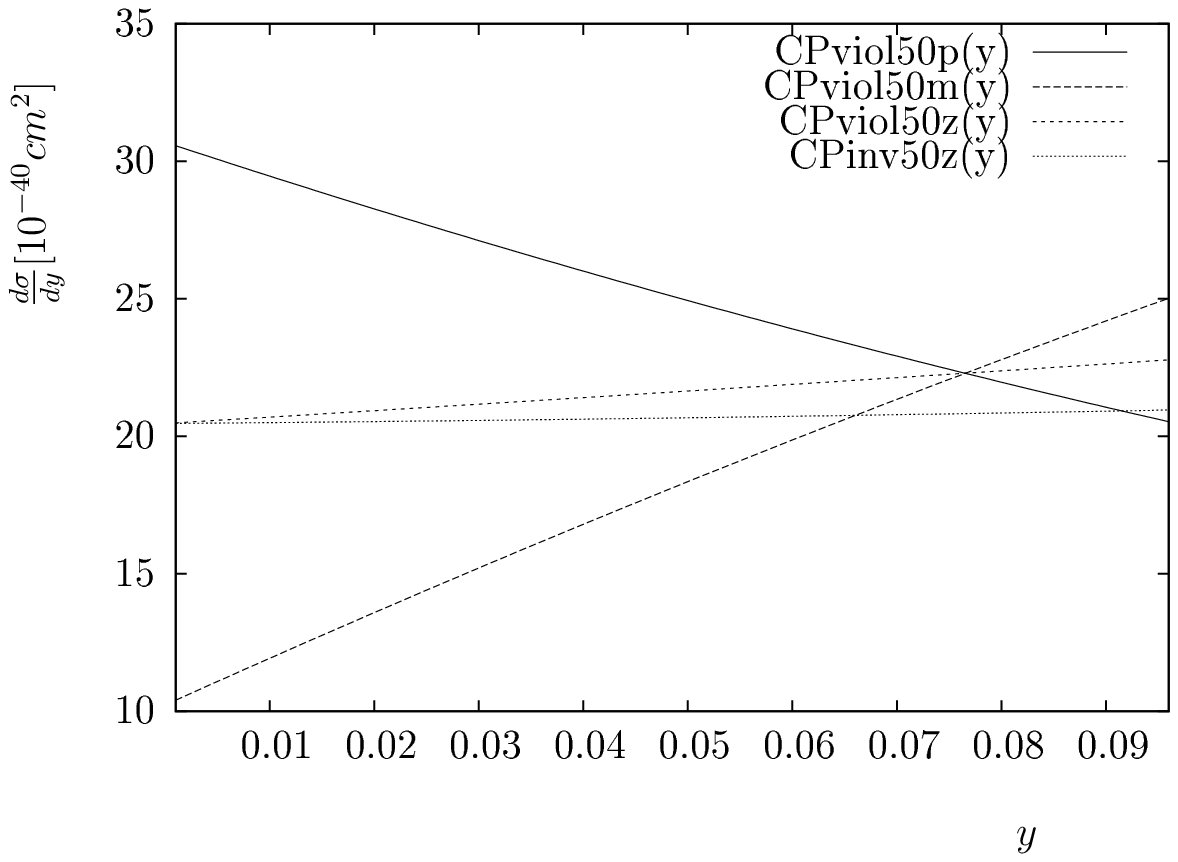}
\end{center}
\caption{Plot of the $\frac{d \sigma}{ d y }_{(V A)}$ as a
function of  $y$ for the
$(\nu_{\mu}p)$ scattering with  $E_{\nu} = 50 \mbox{ MeV}$ and $\mbox{\boldmath$\hat{\eta}_{\nu}$}\cdot\hat{\bf q} = - 1$: 
a) $\mbox{\boldmath $\hat{\eta}_{p}$}\cdot{\bf\hat{q}}=1, |g_A^L| = 1.26,  |g_V^L| = 1, \beta_{VA}=\frac{\pi}{2}$ (solid line), b) $\mbox{\boldmath $\hat{\eta}_{p}$}\cdot{\bf\hat{q}}= -1, |g_A^L| = 1.26,  |g_V^L| = 1, \beta_{VA}=\frac{\pi}{2}$ 
 (long-dashed line),  
c) $\mbox{\boldmath
$\hat{\eta}_{p}$}\cdot{\bf\hat{q}}=0, |g_A^L| = 1.26,  |g_V^L| = 1, \beta_{VA}=\frac{\pi}{2}$ (short-dashed line), d) SM case with 
$\mbox{\boldmath $\hat{\eta}_{p}$}\cdot{\bf\hat{q}}=0, |g_V^L| =1, 
|g_A^L| = 1.26,  \beta_{VA} = \pi $ (dotted line).}\label{spCPv50}
\end{figure}
\begin{figure}
\begin{center}
\includegraphics*[scale=.7]{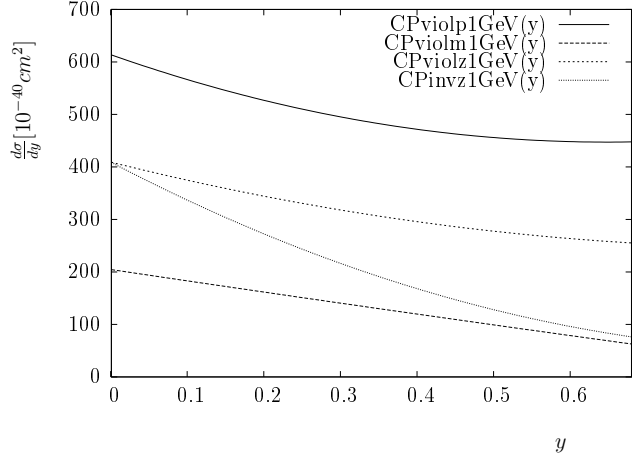}
\end{center}
\caption{Plot of the $\frac{d \sigma}{ d y }_{(V A)}$ as a
function of  $y$ for the
$(\nu_{\mu}p)$ scattering with  $E_{\nu} = 1 \mbox{ GeV}$ and $\mbox{\boldmath$\hat{\eta}_{\nu}$}\cdot\hat{\bf q} = - 1$: 
a) 
$\mbox{\boldmath $\hat{\eta}_{p}$}\cdot{\bf\hat{q}}=1, |g_A^L| = 1.26,  |g_V^L| = 1, \beta_{VA}=\frac{\pi}{2}$ (solid line), b) 
$\mbox{\boldmath $\hat{\eta}_{p}$}\cdot{\bf\hat{q}}=-1, |g_A^L| = 1.26,  |g_V^L| = 1, \beta_{VA}=\frac{\pi}{2}$  
 (long-dashed line), c) 
$\mbox{\boldmath $\hat{\eta}_{p}$}\cdot{\bf\hat{q}}=0, |g_A^L| = 1.26,  |g_V^L| = 1, \beta_{VA}=\frac{\pi}{2}$ (short-dashed line), d) SM case with $\mbox{\boldmath $\hat{\eta}_{p}$}\cdot{\bf\hat{q}}=0, |g_V^L| =1, 
|g_A^L| = 1.26,  \beta_{VA} = \pi$ (dotted line).}  \label{spCPv1GeV}
\end{figure}

\section{Conclusions}

We have shown that  the scattering of the left-chirality muon neutrinos on the PPT can be
used to measure the CP violation in the neutral current semileptonic weak interaction. Admittance of the complex  coupling constants 
$g_V^L, g_A^L$ generates the
interference terms between the standard $g_{V, A}^{L}$ couplings, which are proportional to the $|\mbox{\boldmath $\eta_{p}^{\perp}$}|$ and dependent on the azimuthal angle of the outgoing proton momentum. 
 The azimuthal asymmetry of the recoil protons is independent of 
 the neutrino mass and does not vanish even if
$\beta_{VA} = 0$. The CP-breaking phase $\beta_{VA}$ could be
detected by measuring the maximal asymmetry of the differential cross section. \\
We have also demonstrated that the measurement of the proton energy spectrum in the case of the PPT can be useful  in distinguishing the detector background from the neutrino interactions. \\
To make these tests feasible, the polarized proton target ($10^{30}$ polarized protons and more) and intense neutrino beam ($10^{20}$ neutrinos per year) should be identified. The searching for the CP-violating effects requires the low-threshold, real-time detectors  measuring both  the polar angle and azimuthal angle of the outgoing proton momentum with a high resolution. It is worthwhile mentioning  the proton and neutron polarized targets for nucleon-nucleon experiments at SATURNE II \cite{Saturne}. New low-threshold technology is being developed, e. g.  the silicon cryogenic detectors  and the high purity germanium detectors \cite{nmm}. \\
The experiments using the PPT in the neutrino-nucleon scattering will be a real challenge for experimental groups, but they could detect the existence of the  CP-violating phases and non-standard  neutral and charged current weak interactions. 
In a separate paper, we will search for the exotic effects beyond the SM in the neutrino beam scattering off the polarized nucleon target. 


\begin{thebibliography}{99}

\bibitem{Glashow} S. L. Glashow,   Nucl. Phys.   {\bf 22}  (1961) 579.
\bibitem{Wein} S. Weinberg,   Phys. Rev. Lett.  {\bf 19}  (1967)  1264.
\bibitem{Salam} A. Salam, in  Elementary Particle Theory, (Almquist and Wiksells, Stockholm, 1969).
\bibitem{Balantekin} A. B. Balantekin and G. M. Fuller, J. Phys. G {\bf 29} (2003) 2513. 
\bibitem{Raffert} M. Raffert, H. T. Janka, K. Takahashi and G. Schafer, Astron. Astrophys. {\bf 319} (1999) 122. 
\bibitem{Kneller}  J. P. Kneller, G. C. McLaughlin and R. Surman, J. Phys. G: Nucl. Part. Phys. {\bf 32} (2006) 443. 
\bibitem{Vogel} P. Vogel and J. F. Beacom, Phys. Rev. D {\bf 60} (1999) 053003. 
\bibitem{Beacom} J. F. Beacom, W. M. Farr and P. Vogel,  Phys. Rev. D {\bf 66} (2002) 033001.
\bibitem{Jesus} A. B. Balantekin, J. H. de Jezus, R. Lazauskas and C. Volpe, hep-ph/0603078
\bibitem{Avignone} F. T. Avignone and Y. V. Efremenko, J. Phys. G. {\bf 29} (2003) 2615. 
\bibitem{Scholberg}  K. Scholberg,   Phys. Rev. D {\bf 66} (2002). 
\bibitem{CVolpe} C. Volpe J. Phys. G {\bf 30} (2004) L1. 
 \bibitem{Geer} S. Geer, Phys. Rev. {\bf D 57} (1998) 6989; Phys. Rev. {\bf D 59} (1999) 039903.
\bibitem{Serreau} J. Serreau  and C. Volpe, Phys. Rev. C {\bf 70} (2004) 055502. 
\bibitem{Mc} G. C. McLaughlin, Phys. Rev. C {\bf 73} (2006) 033005. 
\bibitem{Volpe} C. Volpe J. Phys. G {\bf 31} (2005) 903. 
\bibitem{McLau} G. C. McLaughlin and C. Volpe, Phys. Lett. B {\bf 591} 2004 (229). 
\bibitem{JHJesus} A. B. Balantekin, J. H. de Jezus and C. Volpe, Phys. Lett. B {\bf 634} (2006) 180. 
\bibitem{barion} A. Riotto and M. Trodden, Annu. Rev. Nucl. Part. Sci. {\bf 49} (1999) 35.
\bibitem{Kobayashi} M. Kobayashi and T. Maskawa, Prog. Theor. Phys. {\bf 49} (1973)  652.
\bibitem{CP} J.H. Christenson, J.W. Cronin, V.L. Fitch and R. Turlay, 
Phys. Rev. Lett. {\bf 13} (1964) 138; B. Aubert et al., Phys. Rev. Lett.
{\bf 87} (2001) 091801; K. Abe et al., Phys. Rev. Lett. {\bf 87} (2001) 091802.
\bibitem{Todd} CPLEAR Collaboration, A. Angelopoulos et al., Phys.
Lett. B 444 (1998) 43.
\bibitem{neutron} L. J. Lising et al.,  Phys. Rev. C {\bf 62} (2000) 055501.
\bibitem{Saturne} J. Ball et al., Nucl. Instr. and Meth. in Phys. Res. {\bf A} 381 (1996) 4.
\bibitem{Greiner} W. Greiner, B. Muller, Gauge Theory of Weak Interactions, Springer, 2000. 
 \bibitem{Mulan} D. M. Webber et al.,  Phys. Rev. Lett. {\bf 106} (2011) 041803.
 \bibitem{nmm} B.S. Neganov et al., hep-ex/0105083.


\end{thebibliography}
\end{document}